\documentclass{elsart}
\usepackage{graphics}
\usepackage{graphicx}
\usepackage{epsfig}
\usepackage{amssymb,amssymb,rotating}

\begin{document}
\begin{frontmatter}
\title{Clifford wavelets for fetal ECG extraction}
\author{Malika Jallouli}
\address{Universit\'e de Sousse, Ecole Nationale d'Ingénieurs de Sousse, LATIS Laboratory of Advanced Technology and Intelligent Systems, 4023 Sousse, Tunisia.}
\ead{jallouli.malika3@gmail.com}
\author{Sabrine Arfaoui}
\address{Laboratory of Algebra, Number Theory and Nonlinear Analysis LR18ES15, Department of Mathematics, Faculty of Sciences, 5019 Monastir, Tunisia. \\ \& Department of Mathematics, Faculty of Sciences, University of Tabuk, Saudi Arabia.}
\ead{sabrine.arfaoui@issatm.rnu.tn}
\author{Anouar Ben Mabrouk}
\address{Department of Mathematics, Higher Institute of Applied Mathematics and Computer Science, University of Kairouan, Street of Assad Ibn Al-Fourat, Kairouan 3100, Tunisia.\\ \& Laboratory of Algebra, Number Theory and Nonlinear Analysis LR18ES15, Department of Mathematics, Faculty of Sciences, 5019 Monastir, Tunisia. \\ \& Department of Mathematics, Faculty of Sciences, University of Tabuk, Saudi Arabia.}
\ead{anouar.benmabrouk@issatso.rnu.tn}
\author{Carlo Cattani}
\address{Engineering School (DEIM), Tuscia University, Viterbo, Italy.}
\ead{cattani@unitus.it}
%
\begin{abstract}
Analysis of the fetal heart rate during pregnancy is essential for monitoring the proper development of the fetus. Current fetal heart monitoring techniques lack the accuracy in fetal heart rate monitoring and features acquisition, resulting in diagnostic medical issues. The challenge lies in the extraction of the fetal ECG from the mother ECG during pregnancy. This approach has the advantage of being reliable and non-invasive technique. For this aim, we propose in this paper a wavelet/multi-wavelet method allowing to extract perfectly the feta ECG parameters from the abdominal mother ECG. The method is essentially due to the exploitation of Clifford wavelets as recent variants in the field. We prove that these wavelets are more efficient and performing against classical ones. The experimental results are therefore due to two basic classes of wavelets and multi-wavelets. A first class is the classical Haar-Schauder, and a second one is due to Clifford valued wavelets and multi-wavelets. These results showed that wavelets/multiwavelets are already good bases for the FECG processing, provided that Clifford ones are the best.
\end{abstract}
\begin{keyword}
Abdominal ECG; Fetal ECG; wavelets/multiwavelets; Clifford wavelets/multi-wavelets; Haar-Faber-Schauder wavelets/multi-wavelets.\\
\PACS 42C40; 92C55.
\end{keyword}
\end{frontmatter}
\section{{Abbreviations}}
\textbf{ECG} : Electrocardiogram.\\
\textbf{AbdECG} : Abdominal electrocardiogram.\\
\textbf{FECG} : Fetal electrocardiogram.\\
\textbf{MECG} : Mother electrocardiogram. \\
\textbf{FHR} : Fetal Heart Rate.\\
\textbf{EKG} : Electrocardiography.\\
\textbf{WHO} : World Health Organization.\\
\textbf{DAISY} : DAtabase for the Identification of SYstems.\\
\textbf{HFSch} : Haar-Faber-Schauder.\\
\textbf{STFT} : Short time Fourier transform.\\
\textbf{$AFECG_J$} : Approximation of the fetal electrocardiogram at the level $J$.\\
\textbf{$AMECG_J$} : Approximation of the mother electrocardiogram at the level $J$.\\
\textbf{$AS_J$} : Approximation of a signal $S$ at the level $j$. \\
\textbf{$DFECG_J$} : Detail component of the fetal electrocardiogram at the level $J$.\\
\textbf{$DMECG_J$} : Detail component of the mother electrocardiogram at the level $J$.\\
\textbf{$DS_J$} : Detail component of a signal $S$ at the level $J$.\\
\textbf{$DWT_{j,k}$} : Discrete wavelet transform at the level $j$ and the position $k$.
\section{Introduction}
The present paper may be considered are twofold work. One aim is to process some special biosignals by means of wavelets/multiwavelets, and prove that such processors are efficient tools. On the other hand, we aim to show that Clifford wavelets, the most recent forms of wavelets, are more performing and more robust in signal processing compared to the classical most known ones in this field. Clifford wavelets are introduced in harmonic analysis recently. Some initial essays have been developed in some special/simple cases on complex and quaternion spaces. Next, De Bie and Xu, in \cite{DeBie-Xu}, have introduced a famous idea of Clifford-Fourier transform taking into account the angular Gamma operator. This formula permitted to derive an inversion rule, a translation operator and a convolution. These operators are surely the key features behind the signal processing as for the classical Fourier and wavelet methods. A series of works have been next developed by generalizing the classical harmonic analysis such as orthogonal polynomials, Fourier and wavelets to the case of Clifford framework.

In the present paper, we focus on the so-called ECG signals. Recall that cardiovascular disease is the most common cause of the death in the world according to annual WHO statistics. Therefore, the diagnosis of these dangerous diseases is always a vital task. In hospitals' cardiology departments, the electrocardiogram signal remains one of the predominant and most widely used tools for the diagnosis and analysis of cardiac arrhythmia.

In reality, ECG examination is a non-invasive tool performed by bio-physicians to explore the functioning of the heart by the use of external electrodes brought into contact with the skin. It is a signal that reflects the electrical activity of the heart. It informs us about how the heart works by measuring its electrical activity. In fact, with each heartbeat, an electrical impulse (or "wave") passes through the heart. This wave causes a contraction of the heart muscle so that it expels blood from the heart. The ECG measures and records the electrical activity that passes through the heart permitting next to decide whether the electrical activity observed is normal or abnormal.

Although ECG examination is painless and non-invasive, its interpretation remains complex, and requires methodical analysis and some clinical experience. It allows to highlight various cardiac anomalies and has an important place in diagnostic examinations in cardiology, as for coronary artery disease.

On the other hand, the FECG signal reflects the electrophysiological activity of the fetal heart. Congenital heart defects originate in early stages of pregnancy when the heart is forming and they can affect any of the parts or functions of the heart. Cardiac anomalies may occur due to a genetic syndrome, inherited disorder, or environmental factors such as infections or drug misuse \cite{AAmini,FJamali,ARobert,JESchneider}. Fetal abnormalities may be detected during fetal development in time by analyzing the fetal ECG waveform.

FECG is a crucial clinical issue for monitoring the development and well-being of the fetus, throughout pregnancy and childbirth. The challenge is to be able to reliably extract, from external and non-invasive sensors positioned on the mother’s abdomen, an FECG signal of sufficient quality to allow clinical diagnosis. The main difficulty lies in the fact that the abdominal ECG signal of a pregnant woman is a mixture of several signals (MECG, FECG and noise due to uterine contractions and artefacts by movements of the fetus and the mother ...) and that the FECG is of lower energy compared to other present signals.

In this paper, we propose a wavelet/multi-wavelet method permitting to extract the FECG parameters from the MECG. The proposed approach is based on the extraction of significant parameters from the MECG signal reconstructed by suitable wavelets/multi-wavelets. From the reconstructed signal, we manage to eliminate the existing forms of noise and to detect the parameters related to the FECG.

Wavelet analysis appeared in the early 1980s as a multidisciplinary tool that brought together engineers, mathematicians and physicists. The mathematical synthesis led to new results, which brought broader perspectives in each original discipline. By this time, most scientific researchers have heard of wavelets.

Wavelets originated when certain subjects of study required frequency and time analysis simultaneously. In the nineteenth century, Fourier analysis was the only technique allowing the decomposition of a signal into frequencies' components. Unfortunately it provides a frequency analysis but does not allow the temporal localization especially for abrupt changes.

Fourier analysis is based on the fact that functions showing periodicity and certain degree of regularity can be represented by a linear combination of sines and cosines. The coefficients of this linear combination provide information on the level of the frequencies present in the signal.  

The ability to estimate the frequency spectrum of signals as a function of time makes it useful in some cases of ECG processing. Indeed, in medicine, the ECG of a sick patient is different from that of a healthy one. This difference is sometimes very difficult to spot when the EKG is given as a function of time. It becomes evident when it is given as a function of frequencies. The inconvenient is that the Fourier series gives the quantity of each frequency present in the signal for the whole observation period. Fourier theory therefore becomes ineffective for a signal whose frequency spectrum varies considerably over time. Unlike the Fourier analysis, wavelet analysis offers a wide range of basic functions from which one can choose the most appropriate for a given application.

One aim in the present work is to prove that Wavelets may be a successful machinery to conduct applications using a step forward extension of wavelets to multi-wavelets by developing an efficient procedure permitting to extract The FECG from the MECG accurately.

Multi-wavelets have been introduced since the early 1990s as another view of wavelets permitting to re-write wavelet analysis in a vector form. The majority of cases of existing multi-wavelets' constructions, especially in experimental cases, starts from one wavelet or scaling function $\psi/\varphi$ and consider the vector $\Psi=(\psi(.),\psi(.-1),\dots,\psi(.-N))$ or $\Phi=(\varphi(.),\varphi(.-1),\dots,\varphi(.-N))$ where $N$ is the corresponding filter length associated to such functions. This view of wavelets has even though some advantages such as short supports, smoothness, accuracy, symmetry and orthogonality. However, it surely induce some correlation between the components of multi-wavelet decomposition of signals due to the non independence of the multi-wavelet components, especially in non orthogonal case. In the present paper, we will apply differently some types of multi-wavelets where the components are issued each one from a different source. One of them has been already applied in \cite{Zemni2} and has shown to be powerful in estimating biomedical signals. A second variant is due to Clifford wavelets recently constructed in \cite{Arfaoui1,Arfaoui2}. We will show that Clifford wavelets induce in a natural way a variant of multi-wavelets by considering their Clifford components such as the real parts, the vector parts, the bi-vector parts, ..., as wavelets and merge them to obtain a multi-wavelet.

This paper is organized as follows. Section 3 is a brief state of the art of the most common FECG extraction methods. Section 4 is concerned with wavelets and multi-wavelets presentation. We recall the basic steps in construction the wavelets/multi-wavelets to be applied in the present work, such as Haar and Faber-Schauder wavelets and their associated multi-wavelet, and the Clifford wavelets and their associated multi-wavelets. Section 5 is devoted to the development of the bio-experimentation due to the wavelet/multi-wavelet processing of ECG signals in order to extract the FECG from the MECG. The experiments proved the effectiveness of the proposed multi-wavelet theory for extracting the fetal ECG signal in section 5. Besides, they showed the superiority of Clifford wavelets/multi-wavelets as recent variants in wavelet theory. Section 6 is a concluding part, in which we review briefly the results developed in our present work and raise some possible future directions.
\section{FECG extraction brief review}
The FECG which is believed to contain more information than conventional ultrasound methods is always measured by electrodes on the mother’s abdomen. However, the recorded signal suffers always from the mixture of several sources of noise and interference including the very high level of the MECG. In previous studies, several methods have been proposed for extracting the ECG from signals recorded by electrodes placed on the surface of the mother’s body. Despite technological improvements, extracting FECG from abdominal recordings is still a difficult problem that has been addressed by a large number of studies. However, due to the low signal-to-noise ratio of these signals, the application of FECG was limited to the analysis of heartbeats and invasive ECG recordings during childbirth. 

In the present research, the objective is to improve the signal processing methods used in fetal cardiographs, and to provide efficient solutions to this problem, by developing suitable techniques for extracting and filtering ECG signals from the fetuses recorded by an array of electrodes placed on the mother’s womb. So for a better extraction of ECG wave-forms from the fetus in order to aid in the medical diagnosis of cardiac pathology, the approach envisaged consists in improving the estimation of the FECG signal using two wavelet/multi-wavelet based methods such as the one developed in \cite{Zemni1} and consisting of the simplest wavelet/multi-wavelet tollkit and the last recent one developed in \cite{Arfaoui1,Arfaoui2} due to Clifford wavelets as the most recent forms in the field.

In \cite{LIsuu}, the authors proposed to extract the fetal electrocardiogram from a single-lead maternal abdominal ECG. The algorithm is composed of three components. First, the maternal and fetal heart rates are estimated by the de-shaped short time Fourier transform, which is a recently proposed nonlinear time-frequency analysis technique. The beat tracking technique is the second component which is applied to accurately obtain the maternal and fetal R peaks. The third component consists of establishing the maternal and fetal ECG wave-forms by the non local median.

The authors in \cite{Mnikn} presented an extended nonlinear Bayesian filtering procedure for extracting ECG from a single channel as encountered in the fetal ECG extraction from abdominal sensor. The recorded signals are modeled as the summation of several ECG signals. Each of them is described by a nonlinear dynamic model.
\section{Two wavelet/multi-wavelet processors}
In this section, we recall the principal tool in our study consisting of wavelets and their extension to multi-wavelets. 

We proposed in a first step to improve wavelet processing by applying recent families of multi-wavelets issued from single ones where independent components for multi-scaling and multi-wavelet mother functions are used. We will consider as in \cite{Arfaoui1,Arfaoui2,Zemni1,Zemni2} vector-valued mother multi-wavelet 
$\Psi_{HFSch}=\left(\psi_H,\psi_{FSch}\right)$ for the case of Haar-Faber-Schauder multiwavelet essentially issued from \cite{Zemni1}, and $\Psi_{Cl}=\left(\psi_1,\psi_2\right)$ for the case of Clifford multi-wavelets due to \cite{Arfaoui1}.
\subsection{The Haar-Faber-Schauder system}
Recall that Haar mother wavelet ($\psi_H=\chi_{[0,1/2[}-\chi_{[1/2,1[}$) is the most simple case in explicit wavelets. It resembles to piece-wise constant signals, and it has been shown to cover many situations in signal processing. It is compactly supported, not enough regular, explicit, oscillating with one vanishing moment. It yields an orthonormal system $(\psi_H^{j,k})_{j,k\in\mathbb{Z}}$, where $\psi_H^{j,k}(t)=2^{-j/2}\psi_H(2^jt-k)$. More importantly, it is simple to implement. It is adapted more to piece-wise constant (may be periodic) signals.

However, this system may not be well adapted to approximate more complex cases such as piece-wise linear ones for example. In this case, better systems may be adapted. The second system known in functional approximation is the piece-wise linear Faber-Schauder wavelet system based on the mother wavelet 
$2\psi_{FSch}(x)=\Lambda(2x)-2\Lambda(2x-1)+\Lambda(2x-2)$, 
where $\Lambda(x)=\max(0,1-|x|)$. Such a system has been also proved to be suitable in many situations in signal/image processing (See for example \cite{Douzi}). In image processing, like Haar system, the Faber-Schauder wavelet also presents many advantages and important features. Firstly, it possess also an explicit formulation very easy to handle. It is also compactly supported. Moreover, permits the preservation of pixel values range and edge detection. 

These advantages have been encouraging motivations and causes behind \cite{Zemni1} where the authors have developed an entropy based procedure for approximating signals with such wavelets by considering a multi-wavelet case its components are exactly Haar and Faber-Schauder wavelets. In the present work, we continue to exploit such case and consider the Haar-Faber-Schauder multi-wavelet $\Psi_{HFSch}=\left(\psi_{H}\; \psi_{FSch}\right)^T$, where the upper script $^T$ stands for the transpose. This multi-wavelets merges the characteristics of both Haar and Faber-Schauder systems and thus constitutes a better loop for the processing of signals/images. It is indeed compactly supported, explicit, has a reduced number of nonzero recursion coefficients, obtained by recursively averaging and differentiating coefficients.
\subsection{Clifford wavelets and multi-wavelets}
In this subsection we recall briefly the concept of Clifford-valued wavelets and multi-wavelets constructed on the real Clifford algebra $\mathbb{R}_3$ and the useful tools for the associated wavelet analysis to be applied later. Consider the Euclidean space $\mathbb{R}^3$ with its canonical basis $\mathcal{B}=(i,j,k)$, and equipped with an interior product defined on the basis by
$$
i^2=j^2=k^2=-1\quad\hbox{and}\quad
ij+ji=ik+ki=jk+kj=0.
$$
Denote next 
$$
e_1=ij,\;e_2=ik,\;e_3=jk,\;\hbox{and}\;e_4=ijk.
$$
The real Clifford algebra $\mathbb{R}_3$ is the $\mathbb{R}$-algebra with dimension $8$ whom basis is $\widetilde{\mathcal{B}}=(1,i,j,k,e_1,e_2,e_3,e_4)$. Any element $u\in\mathbb{R}_{3}$ is written as
$$
u=\underbrace{u_0}_{\hbox{real part}}+\underbrace{u_1i+u_2j+u_3k}_{\hbox{vector part}}+\underbrace{v_1e_1+v_2e_2+v_3e_3}_{\hbox{bivector part}}+\underbrace{v_4e_4}_{\hbox{trivector part}}.
$$
In the sequel we also need to apply a conjugation rule defined by
$$
\overline{u}=u_0-u_1i-u_2j-u_3k-v_1e_1-v_2e_2-v_3e_3+v_4e_4.
$$
On the Clifford algebra $\mathbb{R}_3$, a function $f:\mathbb{R}^{3}\longrightarrow\mathbb{R}_{3}$ will be expressed as
$$
f(x)=f_0(x)+f_1(x)i+f_2(x)j+f_3(x)k+\widetilde{f}_1(x)e_1+\widetilde{f}_2(x)e_2+\widetilde{f}_3(x)e_3+\widetilde{f}_4(x)e_4,
$$
where the $f_l$ and the $\widetilde{f}_l$, $l=0,1,2,3$ are real-valued functions $\mathbb{R}^{3}$. 

One of the concepts used to construct wavelets on the real Clifford algebra $\mathbb{R}_{3}$ is the notion of monogenicity, based on the Dirac operator 
$$
\partial_{x}=\partial_{x_1}i+\partial_{x_2}j+\partial_{x_3}k.
$$
and the Cauchy-Kowalevski extension (CK-extension). A function $f=f(x_1,x_2,x_3)$ is said to be monogenic on $\mathbb{R}^3$ if $\partial_{x}f=0$. The CK-extension permits to extend $f$ to a Clifford-valued function on $\mathbb{R}^{4}$ by 
\begin{equation}\label{ck-formal-solution}
F(x_0,x)=\exp(-x_0\partial_{x}) f(x)=\displaystyle\sum_{k=0}^{\infty} \displaystyle\frac{(-x_0)^k}{k!}
\partial_{x}^kf(x).
\end{equation}
Exploiting the fact that $F$ is monogenic, we construct Clifford-valued wavelets. One motivation is due to the fact that Clifford wavelets are the last variants of wavelet functions developed by researchers in order to overcome many problems that are not well investigated by classical transforms. The challenging in such concepts is not the wavelet functions themselves but also the structure of Clifford algebras and their flexibility to include different forms of vector analysis in the same time. There are in the literature two main methods to construct Clifford wavelets. The first one is based on Spin groups and thus includes the factor of rotations in the wavelet analysis provided with the translation and dilatation factors. See \cite{Antoine1,Antoine2}. The second is based on monogenic polynomials. These ones constitute natural extensions of orthogonal polynomials to the case of Clifford algebras. Recall that orthogonal polynomials are widely applied in wavelet theory and signal/image processing. See for example \cite{Antoine1,Alvarez-Sansigre}.

In the present work we will serve of the construction conducted in \cite{Arfaoui1,Arfaoui2} where a class of Clifford-Hermite-Jacobi wavelet functions have been introduced by considering the Clifford-weight
$$
\omega_{\alpha,\beta}(u)=(1+|u|^2)^\alpha e^{-\beta|u|^2}.
$$
This leads to a Clifford mother wavelet 
$$
\psi_{\ell}^{\alpha,\beta}(u)=P_{\ell,m}^{\alpha+\ell,\beta+\ell}(u)\omega_{\alpha,\beta}(u),
$$
where the $P_{\ell}^{\alpha,\beta}(u)$ are the Clifford polynomials generated from the CK-extension (\ref{ck-formal-solution}) of $\omega_{\alpha,\beta}$, and which may be expressed as
$$
F^*(t,u)=\displaystyle\sum\limits_{\ell=0}^{\infty}\frac{t^\ell}{\ell!}P_{\ell}^{\alpha,\beta}(u)\,\omega_{\alpha-\ell,\beta-\ell}(u).
$$
By fixing $\alpha=1.5$ and $\beta=\alpha-1$, we obtained the mother Clifford wavelets 
$$
\psi_{1}(\underline{x})=e_1C_1(-2t+t^3)(1+t^2)^{3/2}e^{-t^2/2},
$$
$$
\psi_{2}(\underline{x})=C_2e_1C_3(t+16t^3+24t^5+13t^7+t^9)(1+t^2)^{3/2}e^{-t^2/2},
$$
where the $C_j$'s ($j=1,2$) are normalization constants with respect to the $L^2$-norm. See \cite{Arfaoui1,Arfaoui2} for more details on the original construction of these wavelets. These will be considered as 2-order multi-wavelets by $\psi_{Cl}=\left(\psi_{1}\;\psi_{2}\right)^T$. 
\section{Wavelet and multi-wavelet FECG processing}
In the present section we propose to apply multi-wavelets for the extraction of FECG signal. We will serve from the explicit HFSch multi-wavelets introduced in \cite{Zemni1,Zemni2} as classical classes and the Clifford ones developed recently in \cite{Arfaoui1,Arfaoui2}, and recalled previously as explicit Clifford wavelets.

Each class of the two multi-wavelets has some advantages. The first one is compactly supported, piece-wise linear and permits a reduced number (2 or 3) of nonzero recursion coefficients, sufficient to cover the experiment. The Clifford wavelets/multi-wavelets are highly regular, with Gaussian decay which permits some artificial compactness of the support and thus joins the first one in some characteristics. Moreover, we did not need the computation of the filters coefficients to conduct a multi-wavelet analysis.

{Associated filters such as Gabor and Clifford-Gabor, Hermite and Clifford-Hermite are already developed and proved to be localized in both the spatial and frequency domains. Such localization are basic facts in image/signal processing as they are responsible for the measurement of local structures such as points, lines, edges, and textures in order to facilitate subsequent interpretation of these structures in higher stages (known as high-level vision). More details and facts are developed in \cite{Brackxetal2006,Brackxetal2006a} with applications related to signal processing, image compression, perceptual image quality. See also \cite{Yangetal}.}

{Applying wavelets and/or multiwavelets in the processing resides for a level $J$ of decomposition in a number of positions $k$. When applying a 2-order multiwavelet for example, we get for each level $J$ a first component $A_J^1$ corresponding to an approximation at the level $J$ according to the first component of the 2-order multiwavelet, a second component $A_J^2$ corresponding to an approximation at the level $J$ according to the second component of the 2-order multiwavelet and next a superposition of details components $D_j^1$ and $D_j^2$ ($0\leq j\leq J$) corresponding to the first and the second components of the analyzing multiwavelet respectively. As a result, for the case of 2-order multiwavelet decomposition at a level $J$ we get
\begin{equation}
S_J=A_J^{1}+A_J^{2}+\sum_{j=0}^JD_j^{1}+\sum_{j=0}^JD_j^{2}.
\end{equation}
The task resembles to applying a double (but blind each one against the other) cameras each inducing an independent representation which can be noisy and next superposing these two representations to attenuate the noise resulting from each one and then have a new and final performant image. The operation looks like the phenomenon of installing two surveillance cameras for example to cover the maximum space and thus induce a complete image.}

{To illustrate the closeness of $S_J$ to the original signal $S$, suitable error tolerances will be computed.}

We now describe the multi-wavelet processing of signals. Let $S=\left(S_1,S_2\right)^T$ be 2-dimensional signal. The detail component at a level $J$ of decomposition is 
\begin{equation}\label{DS00}
DS_J=\displaystyle\sum_lD_{J,l}\Psi_{J,l},
\end{equation}
where the multi-wavelet coefficients $D_{J,l}$ are $(2,2)$-matrices. The sum of these detail components induces the approximation of the signal at the level $J$ as
\begin{equation}\label{S00}
AS_J=\displaystyle\sum_{j<J}DS_j.
\end{equation}
As a consequence, the signal $S$ may be approximated at the level $J$ as
\begin{equation}\label{DS00}
S_J=AS_J+DS_J.
\end{equation}
Using (\ref{S00}) the last approximation may be written \begin{equation}\label{S1}
S\simeq DS_J+DS_{J-1}+DS_{J-2}+\dots+DS_0+AS_0.
\end{equation}
Now, the abdominal ECG signal is a compound signal containing both mother own ECG and fetal ECG
\begin{equation}\label{equ1}
AbdECG=MECG+FECG.
\end{equation}
At a decomposition level $J$ we get
\begin{equation}\label{equ2}
AbdECG_J=AMECG_J+DMECG_J+AFECG_J+DFECG_J.
\end{equation}
In ECG processing we know that the MECG signal is widely time-stronger than the FECG signal embedded in it. Moreover, the noises in which the FECG is embedded are also stronger. Therefore, it is naturally that the energy of MECG signal is the highest while the energy of ECG signal is the lowest. This will allow the multi-wavelet approximation coefficients of the decomposed signal to be easily separated and thus the FECG extracted. 

\begin{figure}[!h]
\begin{center}
\includegraphics[scale=0.92]{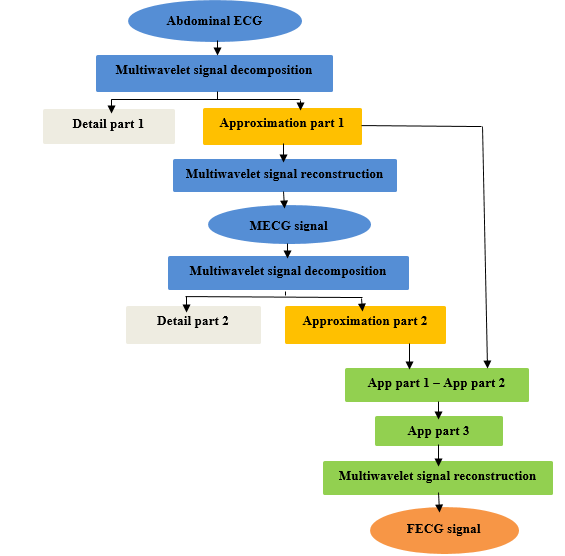}
\caption{The multi-wavelet FECG extraction principle.}
\label{orgFigure32}
\end{center}
\end{figure}

The diagram in Figure \ref{orgFigure32} illustrates the principle of FECG extraction using the multi-wavelet method.

The approximation and detail projections of the FECG signal will be thus extracted as
\begin{equation}\label{equ8}
\left\{\begin{array}{lll}
AFECG_J=AAbdECG_{J}-AMECG_J\\
\hbox{and}\\
AFECG_J=AAbdECG_{J}-AMECG_J.
\end{array}\right.
\end{equation}

Finally, the concept of thresholding and peak detection is used to detect the R-peaks of the FECG signal. An overview of our Method is summarized in Algorithm 1.
\begin{figure}[!h]
\begin{center}
\includegraphics[scale=0.73]{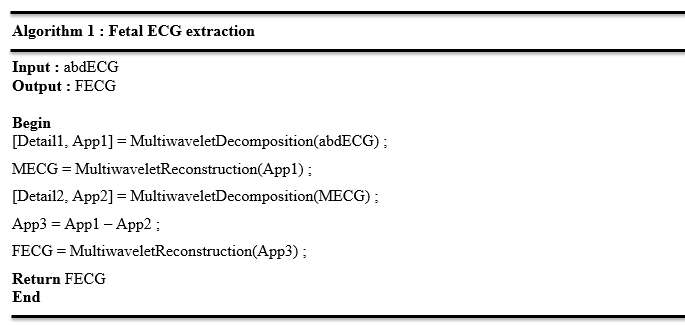}
\label{orgFigure32alg}
\end{center}
\end{figure}

In the experimental part, an abdominal electrocardiogram  signal is applied issued from the DAISY data base. It contains three channels recorded signals for 10 seconds time interval. The proposed method is implemented using MATLAB software. 

\begin{figure}[!h]
\begin{center}
\includegraphics[scale=0.40]{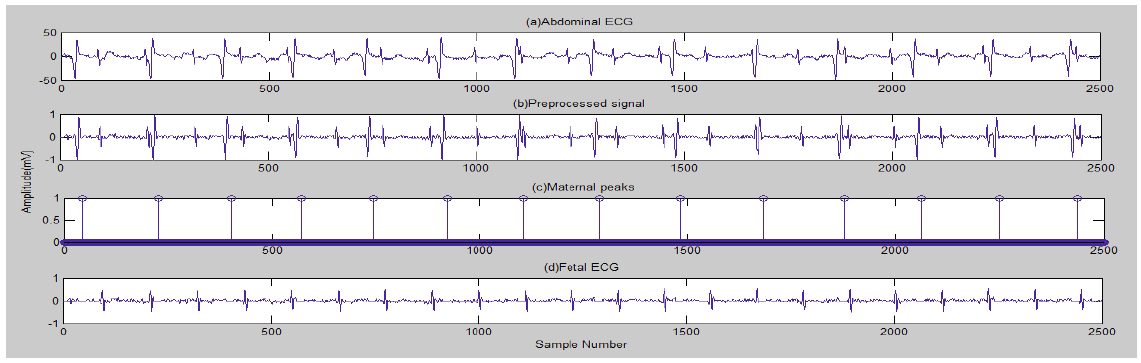}
\caption{Identification of maternal peaks and MECG removal \cite{Book}.}
\label{AnalyseFigure32classic}
\end{center}
\end{figure}

The classical method due to \cite{Book} is implemented using MATLAB software, and is illustrated in Figure \ref{AnalyseFigure32classic}: (a) shows the channel 2 abdECG; (b) pre-processed signal; (c) maternal peaks; and (d) fetal ECG.
The fetal heart rate (FHR) is evaluated as
\begin{equation}\label{EQFHR}
FHR=\frac{Number of peaks detected}{Duration of signal}*60.
\end{equation}
The FHR gives a clear idea of the arrhythmias and other abnormalities in the fetus.

\begin{figure}[!h]
\begin{center}
\includegraphics[scale=0.425]{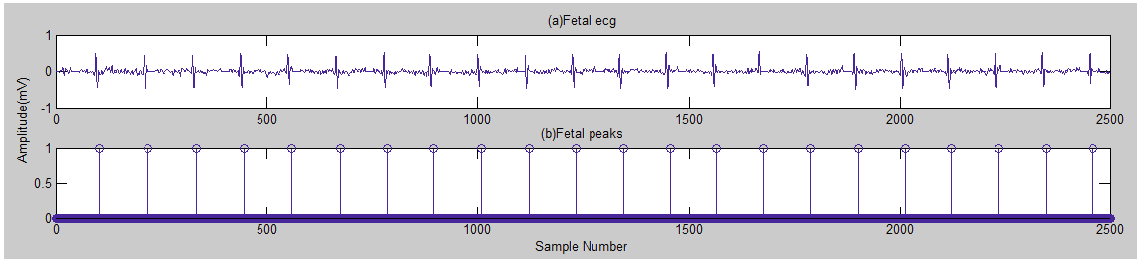}
\caption{The FECG and its detected peaks \cite{Book}.}
\label{AnalyseFigure32classicp}
\end{center}
\end{figure}

The FECG peaks detected are indicated in the Figure \ref{AnalyseFigure32classicp}. A fetal heart rate of 132 bpm (beats per minute) is obtained for channel 2. The normal range of FHR lies between 120 to 160 bpm.

The real peaks which are detected are truly diagnosed (TD) peaks. Some peaks which are detected although they are actually not true are categorized as false positives (FP). An actual peak that is not detected is considered as false negative (FN) \cite{Book}. 

Firstly, to test our method and to evaluate its effectiveness, we implemented it for channels 2.
\begin{figure}[!h]
\begin{center}
\includegraphics[scale=0.70]{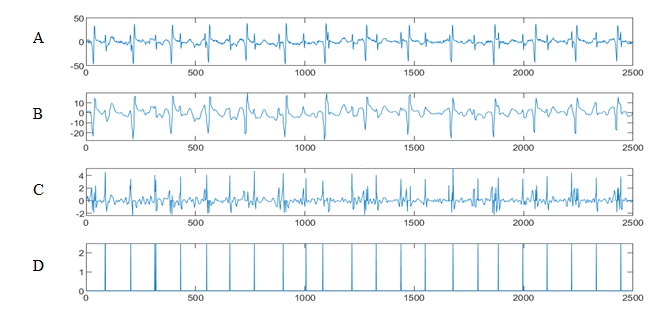}
\caption{FECG extraction and peaks detection using MECGmulti-waveletHFSch MECGmulti-wavelet: (A) AbdECG (B) MECG (C) FECG (D) FECG peaks.}
\label{AnalyseFigure32}
\end{center}
\end{figure}

Figure \ref{AnalyseFigure32} illustrates the result of HFSch multi-wavelet processing. It shows (A) the channel 1 AbdECG, (B) the MECG signal, (C) the FECG signal and (D) the FECG peaks.

\begin{figure}[!h]
\begin{center}
\includegraphics[scale=0.70]{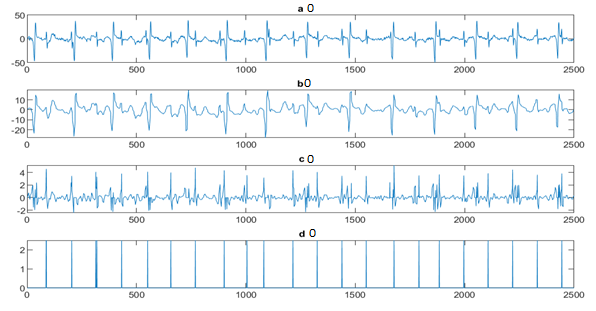}
\caption{FECG extraction and peaks detection using $\psi_0$ clifford MECGmulti-wavelet: (a0) AbdECG (b0) MECG (c0) FECG (d0) FECG peaks.}
\label{AnalyseFigure321}
\end{center}
\end{figure}

\begin{figure}[!h]
\begin{center}
\includegraphics[scale=0.70]{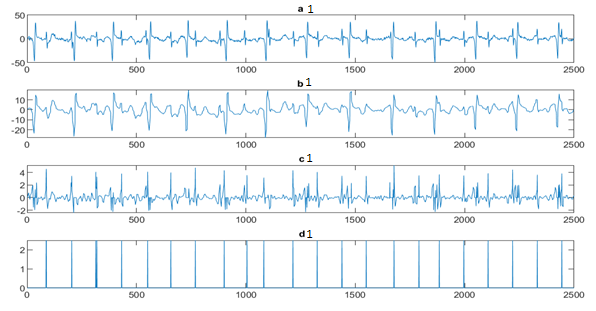}
\caption{FECG extraction and peaks detection  using $\psi_1$ clifford MECGmulti-wavelet: (a1) AbdECG (b1) MECG (c1) FECG (d1) FECG peaks.}
\label{AnalyseFigure322}
\end{center}
\end{figure}

\begin{figure}[!h]
\begin{center}
\includegraphics[scale=0.70]{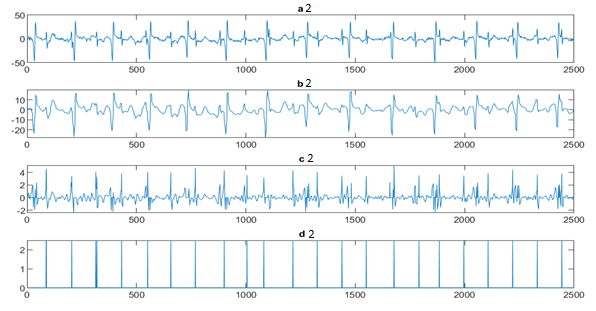}
\caption{FECG extraction and peaks detection  using $\psi_2$ clifford MECGmulti-wavelet: (a2) AbdECG (b2) MECG (c2) FECG (d2) FECG peaks.}
\label{AnalyseFigure323}
\end{center}
\end{figure}

\begin{figure}[!h]
\begin{center}
\includegraphics[scale=0.70]{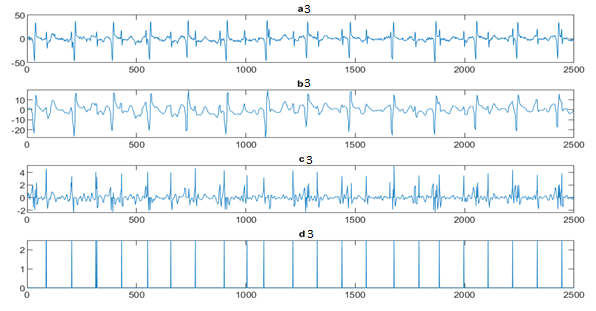}
\caption{FECG extraction and peaks detection  using $\psi_3$ clifford MECGmulti-wavelet: (a3) AbdECG (b3) MECG (c3) FECG (d3) FECG peaks.}
\label{AnalyseFigure324}
\end{center}
\end{figure}

Figures \ref{AnalyseFigure321}, \ref{AnalyseFigure322}, \ref{AnalyseFigure323} and \ref{AnalyseFigure324} illustrate the result of $\psi_0$, $\psi_1$, $\psi_2$ and $\psi_3$ clifford multi-wavelet processing. \\

Next, in order to validate our method and for further assessment, the proposed approach is implemented for channels 3 and 4 of AbdECG. A comparison of the results obtained and those shown in \cite{Book} is summarized in Table \ref{comparisontable1}. It shows the R-peaks detected by our proposed methods and the one in \cite{Book}. It is clear that our approach allows to detect all the peaks present in FECG signal.

The accuracy and sensitivity are estimated and resumed respectively in Table \ref{comparisontable21} and Table \ref{comparisontable22}.
\begin{equation}\label{EQ11}
Accuracy=\frac{TD}{TD+FP+FN}*100.
\end{equation}
\begin{equation}\label{EQ21}
Sensitivity=\frac{TD}{TD+FN}*100.
\end{equation}
\begin{sidewaystable}
\centering
\caption{R-peaks detected.} \label{comparisontable1}
\begin{tabular}{|c|c|c|c|c|c|c|c|}
\hline
Ch. No & Total pks & Pks det in \cite{Book} & MECGmulti-waveletHFSch MECGmulti-wavelet pks det & $\psi_0$ pks det & $\psi_1$ pks det& $\psi_2$ pks det& $\psi_3$ pks det\\
\hline
2&22&22&22&22&22&22&22\\
\hline
3&21&21&21&21&21&21&21\\
\hline
4&21&22&21&21&21&21&21\\
\hline
\end{tabular}
\caption{Accuracy (\%) with different method.} \label{comparisontable21}
\begin{tabular}{|c|c|c|c|c|c|c|}
\hline
Channel number & Acc \cite{Book}  & Acc using MECGmulti-waveletHFSch MECGmulti-wavelet & Acc using $\psi_0$& Acc using $\psi_1$& Acc using $\psi_2$& Acc using $\psi_3$ \\
\hline
2&100&100&100&100&100&100\\
\hline
3&100&100&100&100&100&100\\
\hline
4&86.95&100&100&100&100&100\\
\hline
\end{tabular}
\centering
\caption{Sensitivity (\%) with different method.} \label{comparisontable22}	\begin{tabular}{|c|c|c|c|c|c|c|}	
\hline
Channel number & Sens \cite{Book}  & Sens using MECGmulti-waveletHFSch MECGmulti-wavelet & Sens using $\psi_0$& Sens using $\psi_1$& Sens using $\psi_2$& Sens using $\psi_3$ \\
\hline
2&100&100&100&100&100&100\\
\hline
3&100&100&100&100&100&100\\
\hline
4&95.23&100&100&100&100&100\\
\hline
\end{tabular}
\end{sidewaystable}
Thus our proposed method achieved much better results and all R-peaks of the FECG are detected successfully.

\section{Conclusion} 
In the present paper, wavelet/multi-wavelet processors have been applied for ECG signals processing. Extraction of the FECG signal from the MECG one has been proved to be possible and efficient by using two main sets of wavelets/multi-wavelets such as the Haar-Faber-Schauder system as most recent and simple explicit set, and Clifford wavelets as most newer set of wavelets/multi-wavelets constructed by means of Clifford algebras. 

The experiments proved the effectiveness of the second set in front of the classical example of HFSch, although this set has also proved its efficiency in many cases of signal processing. 

\end{document}